\documentclass{revtex4}

\usepackage{graphicx}

\begin{document}

\title{Discriminating between effective 
theories of $U_{A}(1)$ symmetry breaking}
\author{V. Dmitra\v sinovi\' c}
\affiliation{Research Center for Nuclear Physics 
(RCNP), Osaka University \\ 
Vin\v ca Institute of Nuclear Sciences, P.O.Box 522, Beograd, Yugoslavia}
\begin{abstract}
We address the question if one can empirically distinguish between 
the two proposed solutions to the ``$U_A (1)$ problem'':
the 't Hooft, and the Veneziano-Witten $U_{A}(1)$ 
symmetry breaking effective interactions.
Two hadronic observables are offered as discriminants:
(1) The scalar ($0^{+}$) meson spectrum; 
(2) Weinberg's second spectral sum rule.
Their present experimental status is discussed. 
\end{abstract}
\maketitle

\section{What is the ``$U_A (1)$ problem''?}

The $U_{A}(1)$ problem \cite{nambu67,wein75} consists of two parts:
(i) the discrepancy between the left- and right-hand side in the
``Gell-Mann--Okubo'' type mass relation: 
$$m_{\eta^{'}}^{2} + m_{\eta}^{2} = (1111~ {\rm MeV})^{2} \neq 
2 m_{K}^{2} = (700~ {\rm MeV})^{2},$$
and (ii) the $\eta - \eta^{'}$ mixing angle being negative and far 
from the ideal one:
$$\theta_{\rm ps} \simeq - 20^{o} 
\neq 35.3^{o}~.$$

The presently accepted solution postulates
an explicit breaking of the $U_{A}(1)$ symmetry, i.e., a new 
interaction ${\cal H}_{\rm U(1)}$ that satisfies
\begin{eqnarray}
\left(f m_{\rm U(1)}^{2} f \right)_{ab} 
&=& - \langle 0 | \big[Q_{a}^{5},
[Q_{b}^{5}, {\cal H}_{\rm U(1)}(0)]\big] | 0 \rangle 
\nonumber \\
&=& 0,~~~ a,b = 1, \ldots 8~; 
\nonumber \\
&\neq & 0,~~~ a=b=0 ~~.\
\label{e:dash1}
\end{eqnarray}
Here $a, b$ are the flavour indices of the axial charges
corresponding to the appropriate pseudoscalar (ps) meson(s). 
This interaction raises the
mass of the SU(3) flavour-singlet and thus provides for the 
mass difference 
$$m_{\rm U(1)}^{2} =
m_{\eta^{'}}^{2} + m_{\eta}^{2} - 2 m_{K}^{2} \simeq (855 {\rm MeV})^2 .$$ 
The $U_{A}(1)$ symmetry-breaking mass $m_{\rm U(1)}$ is 855 MeV,
{\rm provided all pseudoscalar decay constants are equal}. 
Inclusion of the variation in ps decay constants
leads to $830 \pm 60~{\rm MeV}$.
The same interaction ${\cal H}_{\rm U(1)}$ solves the ps mixing angle problem:
\begin{eqnarray} 
\tan 2 \theta_{\rm ps} &=&  
{\left(2 \sqrt{2}/ 3\right) \Delta_{\rm ps}^{2} 
\over{\left(1 / 3\right) \Delta_{\rm ps}^{2}
- f_{0}^{2} m_{\rm U(1)}^{2}}} ~, \ 
\label{eigenmassc}
\end{eqnarray}
where
\begin{eqnarray} 
\Delta_{\rm ps}^{2} &=& 
f_{K}^{2} \left(m_{K^{0}}^{2} + m_{K^{+}}^{2}\right)
- f_{\pi}^{2} \left(m_{\pi^{0}}^{2} + m_{\pi^{+}}^{2}\right)
\label{eigenmassa} \\
f_{0}^{2} m_{\rm U(1)}^{2} &=&  
f_{\eta^{'}}^{2} m_{\eta^{'}}^{2} + 
f_{\eta}^{2} m_{\eta}^{2} - f_{K}^{2} \left(m_{K^{+}}^{2} + m_{K^{0}}^{2}
\right) + f_{\pi}^{2} \left(m_{\pi^{+}}^{2} - m_{\pi^{0}}^{2} \right)
\ ~,
\label{eigenmassb}
\end{eqnarray}
which leads to a negative ps. mixing angle
$\theta_{\rm ps} = - (25 \pm 10)^{o}$.

\section{Two solutions}

Two $U_{A}(1)$ symmetry breaking effective operators are discussed
in the literature: 
(1) the't Hooft-Kobayashi-Kondo-Maskawa (``'t Hooft", for short)
effective interaction \cite{th76,kkm71}, 
and (2) the ``Veneziano-Witten" effective interaction 
\cite{ven79,ew79}.

\paragraph*{The 't Hooft interaction}
This interaction is believed to be induced by instantons in QCD. It 
reads
\begin{eqnarray} 
{\cal L}_{\rm tH}^{(N_{f}=3)}  &=& 
- K_{\rm tH} 
\left[ {\rm det}_{f} \left(\bar{\psi} (1 + \gamma_{5}) \psi \right) + 
{\rm det}_{f} \left(\bar{\psi} (1 - \gamma_{5}) \psi \right) \right] 
\label{e:u1a} \
\end{eqnarray}
where, 
${\rm det}_{f}\left(\bar{\psi}(1 \pm \gamma_{5})\psi\right)$
is a determinant of the flavour-space matrix. Eq. (\ref{e:dash1})
then leads to
\begin{eqnarray} 
m^{2}_{00}({\rm tH}) f_{0}^2 &=& 
6 \langle 0 | {\cal L}_{\rm tH}^{(3)}(0) | 0 \rangle + O(1/N_{C})
\nonumber \\
&=& 
- 12 K_{\rm tH} \langle \bar{q} q \rangle_{0}^{3} 
+ O(1/N_{C})~,\
\label{e:th4}
\end{eqnarray}
The symbol $O(1/N_{C})$ remind us that we have 
neglected $1/N_{C}$ suppressed terms.
So long as the respective coupling constant is 
sufficiently large, the U(1) problem is solved. 

\paragraph*{The Veneziano-Witten interaction}
This interaction 
\begin{eqnarray} 
{\cal L}_{\rm VW}^{(N_{f}=3)}  &=& 
K_{\rm VW} 
\left[ {\rm det}_{f} \left(\bar{\psi} (1 + \gamma_{5}) \psi \right) - 
{\rm det}_{f} \left(\bar{\psi} (1 - \gamma_{5}) \psi \right) \right]^{2} 
\label{e:u1b} \
\end{eqnarray}
is {\it not} induced by 
instantons, but rather by $1/N_{C}$ effects in QCD. Again
from Eq. (\ref{e:dash1}) we find
\begin{eqnarray} 
m^{2}_{00}({\rm VW}) f_{0}^2 
&=& 
48 K_{\rm VW} \langle \bar{q} q \rangle_{0}^{6} + O(1/N_{C}) 
~.\
\label{e:vw3}
\end{eqnarray}
The same comments about the coupling constant hold as for the 
't Hooft interaction. Once again, the flavour-singlet pseudoscalar mass 
moves up above the octet one.

\section{Discriminating between the solutions}

Manifestly, no study of the pseudoscalar $\eta$, $\eta^{'}$ mesons'
properties alone can resolve this issue. We offer two new tests 
discriminating between the two effective interactions in a chiral 
quark model. 
Differences between models with the 't Hooft- and the 
Veneziano-Witten (VW) 
$U_{A}(1)$ symmetry-breaking interactions arise in: 
(i) the {\it scalar} mesons spectra, 
\cite{dmitra96,bonn95,dm97}. The 't Hooft interaction 
leads to a mass shift within the scalar nonet that is identical 
in size, but opposite in sign to that found in pseudoscalars, 
whereas the VW one does not shift the scalar meson masses at all.
(ii) the second spectral (Weinberg) sum rule \cite{vd98}. 
The 't Hooft interaction strongly 
modifies this sum rule, whereas the VW one does not change it at all.

\subsection{Scalar meson spectrum}

In the following we use an effective chiral field theory of quarks 
with a non-trivial ground state characterized by a finite
quark condensate and an effective $U_{A}(1)$ symmetry-breaking 
interaction, following Nambu and Jona-Lasinio (NJL) \cite{nambu61}.
The $U_{L}(3) \times U_{R}(3)$ symmetric 
gluon-exchange interaction (the first line) is modelled by a quartic quark
selfinteraction
\begin{eqnarray} 
{\cal L}_{\rm NJL}^{(3)} = 
\bar{\psi} \big[ i {\partial{\mkern -10.mu}{/}} - m^{0} \big]
\psi 
  &+& 
G\sum_{i=0}^{8}
 \big[(\bar{\psi} \mbox{\boldmath$\lambda$}_{i} \psi)^{2} + 
  (\bar{\psi} i \gamma_{5} \mbox{\boldmath$\lambda$}_{i} \psi)^{2} \big]
+ {\cal L}_{\rm U(1)}~,
\label{e:njl2}
\end{eqnarray}
ammended by the $U(1)_{A}$ symmetry breaking effective interaction.
There are at present no easily applicable nonperturbative methods 
for a direct
approach to the 6- or 12-point operators in Eq. (\ref{e:njl2}).
Therefore one has to construct an ``effective mean-field quartic
self-interaction Lagrangian" ${\cal L}_{\rm eff}^{(4)}$ from 
Eq. (\ref{e:u1a},\ref{e:u1b}) following the procedure employed 
in Ref. \cite{dmitra96}. This
leads to consistent chiral dynamics in the sense that the
Goldstone theorem and other chiral Ward-Takahashi identities
pertaining to the ps octet remain intact in the chiral limit. 
Mathematically that procedure is 
equivalent to taking a quark and an antiquark external line 
and closing them into a loop 
using Feynman rules for the Lagrangian (\ref{e:njl2}) in all
possible ways while taking into account the proper symmetry
number of the diagram.
The meson masses are read off from the poles of their propagators,
which in turn are constrained by the gap equation. 
This model has turned out to be a reliable laboratory for calculating 
light spinless meson mass relations induced by $U_{A}(1)$ 
symmetry-breaking, as can be seen from the comparison between 
the NJL model results \cite{dmitra96} and a confining potential
model's predictions \cite{bonn95}. The close agreement of the 
spectra is the best justification of the NJL model.

\paragraph*{'t Hooft interaction}
The ps meson flavour singlet - octet mass shift
due to the 't Hooft interaction in the NJL model has been 
established to be
in exact agreement with the general result (\ref{e:th4}), see 
Ref. \cite{dmitra96}.
One finds, however, that the singlet - octet mass splitting 
in the scalar ($0^{+}$) channel is just as large as, though of 
opposite sign to the ps one. This statement is embodied in
the $N_{f} = 3$ scalar -- pseudoscalar meson mass sum rule 
\begin{eqnarray} 
m^{2}_{\eta} + m^{2}_{\eta^{'}} - m^{2}_{K^{+}} - m^{2}_{K^{0}}  
&=& m^{2}_{K_{0}^{*+}} + m^{2}_{K_{0}^{*0}}  -
m^{2}_{f_{0}} - m^{2}_{f_{0}^{'}}~.
\label{umass4}
\end{eqnarray}
Equivalent results were found in a confining chiral quark model, 
Ref. \cite{bonn95}, also as an effect of the 't Hooft $U_{A}(1)$ 
symmetry breaking interaction. 
For two flavours the `t Hooft interaction predicts a 
mass splitting between the isoscalar ($f_{0}^{*}$)
and isovector ($a_{0}$) scalar ($0^{+}$) states that has been 
observed on the lattice, see Fig. 6 in Ref. \cite{kars00}. The
corresponding pseudoscalar mass splitting with two flavours has 
not been calculated, so direct comparison with the $N_{f} = 2$ 
sum rule 
\begin{eqnarray} 
m^{2}_{\pi} - m^{2}_{\eta^{*}} 
&=& m^{2}_{f_{0}^{*}} - m^{2}_{a_{0}}~
\label{e:nf=2}
\end{eqnarray}
is not possible at the moment.
The sum rule (\ref{umass4}) shifts the masses of the physical
iso-singlet scalar states $f_{0}, f_{0}^{'}$ from their simple
quark model values, see Fig. 1. 
\begin{figure}
\resizebox{25pc}{!}{\includegraphics[width=0.9\textwidth]{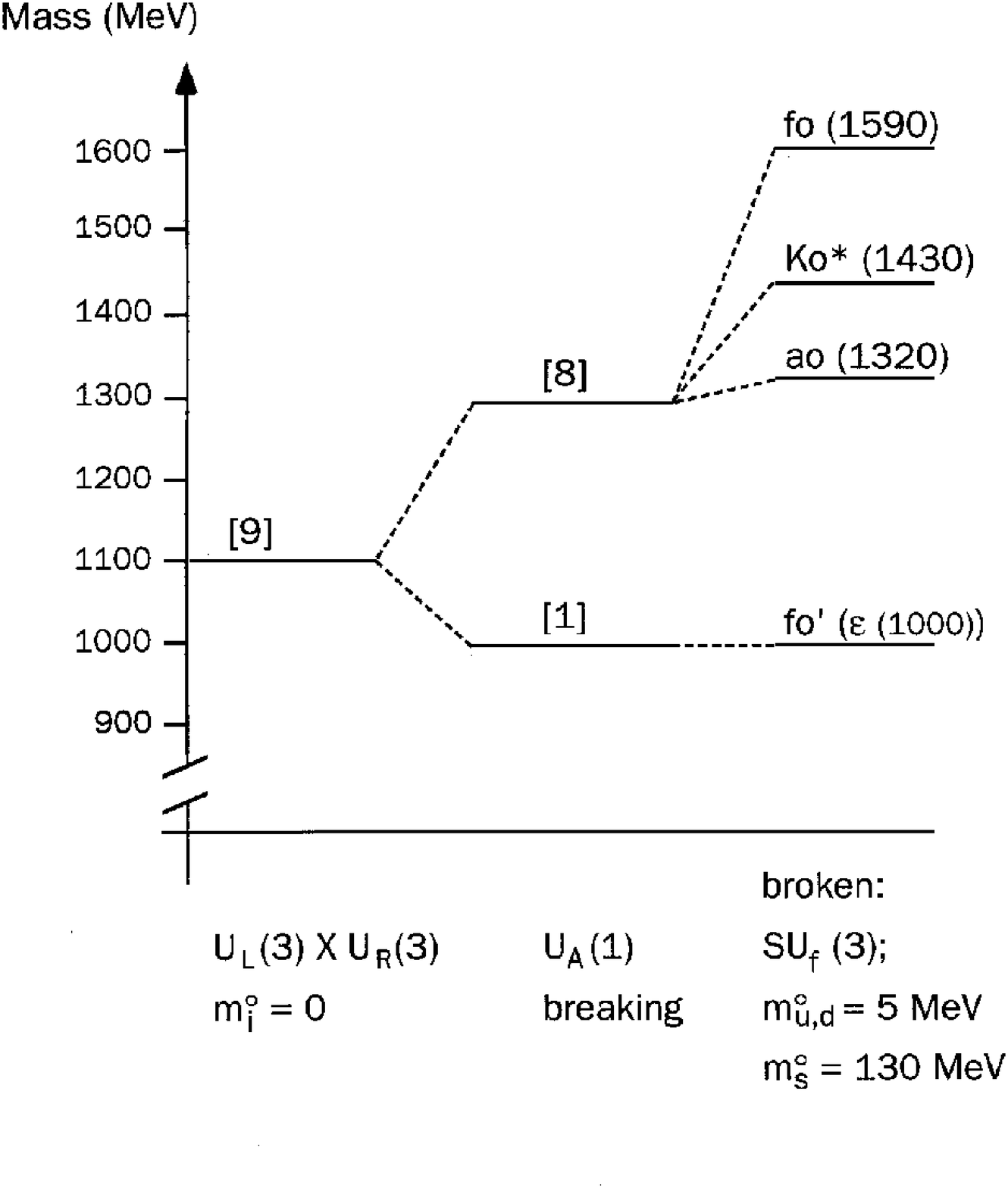}} 
\caption{The scalar meson spectrum in the three flavour NJL model 
with `t Hooft interaction.}
\label{f:scalars}
\end{figure}

Assuming that the $f_{0}(1500)$ is one of the 
two isoscalar scalar states, the sum rule (\ref{umass4}) predicts 
the mass of the other scalar state as 1000$\pm$50 MeV. 
As there are 
{\it two} iso-singlet scalar states $f_{0}$ in the Particle Data 
tables with mass(es) very close to 1 GeV, the
$f_{0}(980)$ and the (``$\sigma$") $f_{0}(\varepsilon (1000))$, 
one is left with an ambiguity. 
The $f_{0}(1500)$ is in better shape:
Ritter et al. \cite{bonn96} have explained the 
puzzling absence of $K \bar K$ pairs from the $f_{0}(1500)$ 
two-body decay products as a consequence of the 't Hooft 
interaction.
This explanation depends critically on the scalar
mixing angle $\theta_{\rm s}$ being small and {\it positive}, 
which follows from
\begin{eqnarray} 
\tan 2 \theta_{\rm s} &=& {\left(4 \sqrt{2}/ 3\right) 
\left(m_{K_{0}^{*}}^{2} - m_{a_{0}}^{2} \right)
\over{m_{\rm U(1)}^{2} + 
\left(2 / 3\right) \left(m_{K_{0}^{*}}^{2} - m_{a_{0}}^{2}
\right)}} ~. 
\label{eigenmassd} \
\end{eqnarray}
One must say that higher-order corrections are known to modify our 
sum rule (\ref{umass4}). 
For example, upon taking into account of vector- 
and axial-vector mesons, the r.h.s. (scalar masses) of Eqs. 
(\ref{umass4}),(\ref{e:nf=2}) are reduced by a multiplicative 
factor equal to the flavour-singlet axial coupling constant of the 
constituent quarks \cite{dm01}.

\paragraph*{Veneziano-Witten interaction}
To leading order in $N_C$ we find again that the general result 
Eq. (\ref{e:vw3}) holds for the 
meson masses, see Ref. \cite{dm97}.
This time scalar mesons are unaffected by the VW interaction. 
In the absence 
of flavour singlet-octet mass splitting in the scalar sector, 
the flavour-singlet scalar mesons mix ideally (see
Eq. (\ref{eigenmassd}), but with $m_{\rm U(1)}^2$ omitted 
from the denominator) and one finds one $u {\bar u}, d {\bar d}$ and 
one $s {\bar s}$ state, with a mass splitting of 
about 300 MeV. 
The lower-lying state is degenerate with the isovector 
scalar mesons, i.e., 
around 1320 MeV in this model.
Curiously, there is an $f_{0}$ state 
at 1370 MeV. Then the heavy scalar meson ought to be near 
1600 MeV. There are two candidates in the vicinity:
(a) the familiar $f_{0}(1500)$, and (b) the new $f_{0}(1720)$.
The former has a puzzling absence, for an $s \bar s$ state, 
of the $K \bar K$ decay mode. 
This has prompted suggestions that it is not an ordinary
$q \bar q$ octet member, as the Veneziano-Witten model predicts.
This evidence and the apparent success of the 't Hooft model
at explaining the $f_{0}(1500)$ decay pattern \cite{bonn96} 
seem to rule against the Veneziano-Witten model.

\subsection{The second spectral sum rule}

Weinberg's second sum rule (Wsr II) \cite{wein67} 
\begin{equation}
\int_{0}^{\infty} ds \left(\rho^{V}(s) - \rho^{A}(s)\right) = 0
\: ,
\label{e:wsr2}
\end{equation}
for the (difference of) vector and axial vector spectral functions 
is a statement about the chiral 
symmetry of the underlying theory at asymptotically large momenta. 
Here $\rho_{V,A}^{ab}$ are the spectral functions as defined by 
\begin{eqnarray}
\langle 0 |\left[A_{\mu}^{a}(x), A_{\nu}^{b}(y)\right]| 0 
\rangle &=& -
i \int d \mu^{2} \rho_{A}^{ab}(\mu^{2}) \left(g_{\mu \nu} - 
{\partial_{\mu}^{x} \partial_{\nu}^{y} \over \mu^{2}} \right) 
\Delta (x - y;\mu^{2}) \nonumber \\
&& + 
i \int d \mu^{2} \rho_{PS}^{ab}(\mu^{2}) 
\partial_{\mu}^{x} \partial_{\nu}^{y}
\Delta (x - y;\mu^{2}) \\
\langle 0 |\left[V_{\mu}^{a}(x), V_{\nu}^{b}(y)\right]
| 0 \rangle &=&
- i \int d \mu^{2} \rho_{V}^{ab}(\mu^{2}) 
\left(g_{\mu \nu} - 
{\partial_{\mu}^{x} \partial_{\nu}^{y} \over \mu^{2}} \right) 
\Delta (x - y;\mu^{2})~\ ,
\label{e:com}
\end{eqnarray}
$V_{\mu}^{a}, A_{\nu}^{b}$ are the vector- and axial currents and 
$\Delta$ is the commutator of the free scalar fields at two space-time 
points. 

The two Weinberg spectral sum rules have been examined in an effective 
field theoretical model of QCD, Ref. \cite{dmit96}, with the result 
that the first sum rule is exactly satisfied, and the second one 
is broken {\it even in the exact $SU_{L}(2) \times SU_{R}(2)$ 
chiral limit}, i.e., with current quark masses
$m^{0}_{u}=m^{0}_{d}=0$.

\paragraph*{Violations of the second Weinberg sum rule}
Nieh \cite{nieh67} gave a critical assessment of 
this sum rule very early, though it passed largely unnoticed.
One can express the violation of the Wsr II due to 
the Hamiltonian $H$ as
\begin{eqnarray}
\delta_{ij} \int_{0}^{\infty} ds 
\left(\rho_{V}^{ab}(s) - \rho_{A}^{ab}(s)\right) 
&=& \int d^{3}x 
\langle 0 |\left[\left[H, A_{i}^{a}(0, {\bf x})\right], 
A_{j}^{b}(0, {\bf y})
\right]| 0 \rangle  
\nonumber \\
&-& \int d^{3}x 
\langle 0 |\left[\left[H, V_{i}^{a}(0, {\bf x})\right], 
V_{j}^{b}(0, {\bf y})\right]| 0 \rangle  ~\ ,
\label{e:nieh}
\end{eqnarray}
where ($i,j=1,2,3$) are the spatial dimension indices.
Thus, Eq. (\ref{e:nieh}) shows that the second sum rule 
actually tests the commutators of the
{\it spatial current components} and the Hamiltonian, i.e., 
the invariance of the theory under $U_{L}(6) \times U_{R}(6)$ 
current algebra symmetry transformations \cite{fgz64}, rather 
than merely the usual $U_{L}(3) \times U_{R}(3)$ chiral charge
algebra, which is its subalgebra. 
Manifestly, any term in the Hamiltonian
that breaks the $U_{L}(3) \times U_{R}(3)$ symmetry will also 
break the $U_{L}(6) \times U_{R}(6)$ symmetry. 
There are three sources of $U_{L}(3) \times U_{R}(3)$ symmetry
breaking in QCD: (i) the current quark masses; 
(ii) $U_{A}(1)$ symmetry-breaking effective interaction; and 
(iii) the electroweak (EW) interactions. 
In the following we shall examine only the first two.

The second spectral sum rule Eq. (\ref{e:nieh}), upon Nieh's correction
\cite{nieh67}, is also satisfied in the effective model of 
Ref. \cite{dmit96}. 
That model already contains the $U_{A}(1)$ symmetry breaking 
't Hooft interaction in its two-flavor version.
Thus there is a large violation of the Wsr II even in the chiral
limit, due the $U_{A}(1)$ symmetry breaking 't Hooft interaction 
in this model \cite{vd98}.

\paragraph*{Current quark mass terms}
Inserting the current quark mass Hamiltonian 
${\cal H}_{\chi SB}(0)={\bar \Psi}(0)M_{q}^{0}\Psi(0)$ into 
Eq. (\ref{e:nieh}) we find
\begin{eqnarray}
\int_{0}^{\infty} ds \left(\rho_{V}^{ab}(s) - \rho_{A}^{ab}(s) 
\right)
&=& \int d^{3}x 
\langle 0 |\left[\left[H_{\chi SB}, A_{i}^{a}(0, {\bf x})\right], 
A_{j}^{b}(0, {\bf y})
\right]| 0 \rangle  
\nonumber \\
&-& \int d^{3}x 
\langle 0 |\left[\left[H_{\chi SB}, V_{i}^{a}(0, {\bf x})\right], 
V_{j}^{b}(0, {\bf y})\right]| 0 \rangle 
\nonumber \\
&=& -  
\langle 0 |{\bar \Psi}
\left\{ \left\{ M_{q}^{0}, {\lambda^a \over 2}
\right\}, \frac{\lambda^b}{2} \right\}\Psi | 0 \rangle
\nonumber \\ 
&=& 
- \langle 0 \left| \left[Q^{a}_{5}, \left[Q^{b}_{5}, 
{\bar \Psi}M_{q}^{0}\Psi \right]\right] \right| 0 \rangle 
\nonumber \\
&=&
\left(f_{\rm ps} m_{\rm ps}^{2}({\rm mech}) f_{\rm ps} \right)_{ab} \ .
\label{e:mass}
\end{eqnarray}
The expression on the right-hand side of (\ref{e:mass}) is
the same as the one entering the Gell-Mann-Oakes-Renner 
(GMOR) formula relating the pseudoscalar (ps) meson masses and decay 
constants to the above vacuum matrix elements. Eq. (\ref{e:mass}) is in 
agreement with the ITEP sum rule results \cite{pasc82}.

\paragraph*{$U_{A}(1)$ symmetry-breaking effective interactions}
Insert the effective 't Hooft quark self-interaction 
Eq. (\ref{e:u1a}) into the double commutators in Eq. (\ref{e:nieh})
to find 
\begin{eqnarray}
\int d^{3}x 
\langle 0 |\left[
\left[H_{\rm tH}^{(3)}, A_{i}^{a}(0, {\bf x})\right], 
A_{j}^{b}(0, {\bf y}) \right]| 0 \rangle  
&-&
\int d^{3}x 
\langle 0 |
\left[\left[H_{\rm tH}^{(3)}, V_{i}^{a}(0, {\bf x})\right], 
V_{j}^{b}(0, {\bf y}) \right]| 0 \rangle 
\nonumber \\
&=& - 6 \langle 0 |{\cal H}_{\rm tH}^{(3)}| 0 \rangle 
\delta_{ij} \delta^{ab}\ .
\label{e:thcom}
\end{eqnarray}
The interacting ground state (``vacuum") expectation value of 
the 't Hooft 
interaction is related to the 't Hooft mass with three light 
flavours {\it via} Eqs. (\ref{e:th4},\ref{e:dash1}), see Ref. 
\cite{dmitra96,dm97}.
The empirical value of $f_{0}^{2} m_{\rm tH}^{2}$ was determined
above as $(300 {\rm MeV})^{4}$ from the 
definition Eq. (\ref{e:th4}). This leads to
\begin{eqnarray}
\int d^{3}x 
\langle 0 |\left[
\left[H_{\rm tH}^{(3)}, A_{i}^{a}(0, {\bf x})\right], 
A_{j}^{b}(0, {\bf y}) \right]| 0 \rangle  
&-&
\int d^{3}x 
\langle 0 |
\left[\left[H_{\rm tH}^{(3)}, V_{i}^{a}(0, {\bf x})\right], 
V_{j}^{b}(0, {\bf y})\right]| 0 \rangle
\nonumber \\
&=& 
\delta_{ij} \delta^{ab} f_{0}^{2} m_{\rm U(1)}^{2} \ .
\label{e:u1}
\end{eqnarray}
Note that this result holds for {\it all} $a, b = 0, ..., 8$,
{\it i.e.}, not only in the flavour-singlet channel ($a, b = 0$).
This is something of a surprise, as we have come to 
expect its influence only in the flavour-singlet ps and scalar
channels. 
Here one is sensitive to the violation of the 
$U_{L}(6) \times U_{R}(6)$ {\it current} algebra, rather than
that of the (usual) 
$SU_{L}(3) \times SU_{R}(3)$ algebra of chiral {\it charges},
and that the 't Hooft interaction violates the $U_{L}(6) \times 
U_{R}(6)$ symmetry.
Adding now the current quark mass contribution to the right-hand side 
of Eq. (\ref{e:u1}) we find
\begin{eqnarray}
\int_{0}^{\infty} ds \left(\rho_{V}^{ab}(s) - \rho_{A}^{ab}(s) 
\right)
&=& 
\delta_{ab} f_{0}^{2} m_{\rm U(1)}^{2} + 
f_{a} m_{ab}^{2}({\rm mech.}) f_{b} \ .
\label{e:tot}
\end{eqnarray}

There is, however, another way to effectively break 
the $U_{A}(1)$ symmetry with quark degrees of freedom: the 
Veneziano-Witten effective quark interaction Eq. (\ref{e:u1b})
Insert this into the double commutators in Eq. (\ref{e:nieh});
direct calculation yields {\it zero}, to leading order in $1/N_{C}$,
\begin{eqnarray}
\int d^{3}x 
\langle 0 |\left[
\left[H_{\rm VW}^{(3)}, A_{i}^{a}(0, {\bf x})\right], 
A_{j}^{b}(0, {\bf y})\right]| 0 \rangle  
&-&
\int d^{3}x 
\langle 0 |
\left[\left[H_{\rm VW}^{(3)}, V_{i}^{a}(0, {\bf x})\right], 
V_{j}^{b}(0, {\bf y})\right]| 0 \rangle
\nonumber \\
&=& 0 + O(1/N_{C})
\ ,
\label{e:vwcom}
\end{eqnarray}
thus leaving only the current quark mass induced corrections to
the Wsr II
\begin{eqnarray}
\int_{0}^{\infty} ds \left(\rho_{V}^{ab}(s) - \rho_{A}^{ab}(s) 
\right)
&=& 
f_{a} m_{ab}^{2}({\rm mech.}) f_{b}  \ ,
\label{e:tot2}
\end{eqnarray}
in the Veneziano-Witten model.
Hence there is an order of magnitude 
difference between these two models of $U_{A}(1)$ symmetry 
breaking in all the flavour channels ($a,b=0,1,\ldots 8$). 

Just one precise measurement, say in the isovector channel,
should discriminate between the two models.
Some data at low energies already exist, see Fig. 1. in Ref. 
\cite{vd98}, but the range of energy integration 
is limited by the $\tau$ lepton mass, and saturation 
of the sum rule is not achieved. 
Thus far, the results are inconclusive. New kinds of experiments seem 
necessary. There is hope, however, that the methods 
described by Hatsuda in these proceedings \cite{hatsuda01} will  
allow an ``exact'' calculation of the second spectral sum rule in
lattice QCD.

\section{Conclusions}

1. There are two effective 
$U_{A}(1)$ symmetry breaking interactions: `t Hooft and Veneziano-Witten.

2. Scalar meson spectrum 
and the second spectral sum rule discriminate between them.

3. Flavour-singlet scalar mesons can be identified in accord with 
either the 't Hooft or Veneziano-Witten interactions. Decay properties 
seem to slightly prefer `t Hooft, but more work is necessary.

4. Present ($\tau$ lepton decay) data on the second Weinberg sum rule 
does not extend high enough in energy to 
differentiate between these two interactions.

\bibliographystyle{apsrev}
\bibliography{kor}

\begin{thebibliography}{10}
\expandafter\ifx\csname bibnamefont\endcsname\relax
  \def\bibnamefont#1{#1}\fi
\expandafter\ifx\csname bibfnamefont\endcsname\relax
  \def\bibfnamefont#1{#1}\fi
\expandafter\ifx\csname url\endcsname\relax
  \def\url#1{\texttt{#1}}\fi
\expandafter\ifx\csname urlprefix\endcsname\relax\def\urlprefix{URL }\fi
\expandafter\ifx\csname bibinfo\endcsname\relax \def\bibinfo#1#2{#2}\fi
\expandafter\ifx\csname eprint\endcsname\relax \def\eprint#1{#1}\fi

\bibitem{nambu67}
\bibinfo{author}{\bibfnamefont{Y.}~\bibnamefont{Nambu}},
  \emph{\bibinfo{title}{Symmetry Principles and Fundamental Particles}}
  (\bibinfo{publisher}{W.H. Freeman and Co.}, \bibinfo{address}{San Francisco},
  \bibinfo{year}{1967}).

\bibitem{wein75}
\bibinfo{author}{\bibfnamefont{S.}~\bibnamefont{Weinberg}},
  \bibinfo{journal}{Phys. Rev. D} \textbf{\bibinfo{volume}{11}},
  \bibinfo{pages}{3583} (\bibinfo{year}{1975}).

\bibitem{th76}
\bibinfo{author}{\bibfnamefont{G.}~\bibnamefont{`t~Hooft}},
  \bibinfo{journal}{Phys. Rev. D} \textbf{\bibinfo{volume}{14}},
  \bibinfo{pages}{3432} (\bibinfo{year}{1976}).

\bibitem{kkm71}
\bibinfo{author}{\bibfnamefont{M.}~\bibnamefont{Kobayashi}},
  \bibinfo{author}{\bibfnamefont{H.}~\bibnamefont{Kondo}}, \bibnamefont{and}
  \bibinfo{author}{\bibfnamefont{T.}~\bibnamefont{Maskawa}},
  \bibinfo{journal}{Prog. Theor. Phys.} \textbf{\bibinfo{volume}{45}},
  \bibinfo{pages}{1955} (\bibinfo{year}{1971}).

\bibitem{ven79}
\bibinfo{author}{\bibfnamefont{G.}~\bibnamefont{Veneziano}},
  \bibinfo{journal}{Nucl. Phys. B} \textbf{\bibinfo{volume}{159}},
  \bibinfo{pages}{213} (\bibinfo{year}{1979}).

\bibitem{ew79}
\bibinfo{author}{\bibfnamefont{E.}~\bibnamefont{Witten}},
  \bibinfo{journal}{Nucl. Phys. B} \textbf{\bibinfo{volume}{156}},
  \bibinfo{pages}{269} (\bibinfo{year}{1979}).

\bibitem{dmitra96}
\bibinfo{author}{\bibfnamefont{V.}~\bibnamefont{Dmitra{\v s}inovi{\' c}}},
  \bibinfo{journal}{Phys. Rev. C} \textbf{\bibinfo{volume}{53}},
  \bibinfo{pages}{1383} (\bibinfo{year}{1996}).

\bibitem{bonn95}
\bibinfo{author}{\bibfnamefont{E.}~\bibnamefont{Klempt}},
  \bibinfo{author}{\bibfnamefont{B.~C.} \bibnamefont{Metsch}},
  \bibinfo{author}{\bibfnamefont{C.~R.} \bibnamefont{M{\" u}nz}},
  \bibnamefont{and} \bibinfo{author}{\bibfnamefont{H.~R.} \bibnamefont{Petry}},
  \bibinfo{journal}{Phys. Lett. B} \textbf{\bibinfo{volume}{361}},
  \bibinfo{pages}{160} (\bibinfo{year}{1995}).

\bibitem{dm97}
\bibinfo{author}{\bibfnamefont{V.}~\bibnamefont{Dmitra{\v s}inovi{\' c}}},
  \bibinfo{journal}{Phys. Rev. D} \textbf{\bibinfo{volume}{56}},
  \bibinfo{pages}{247} (\bibinfo{year}{1997}).

\bibitem{vd98}
\bibinfo{author}{\bibfnamefont{V.}~\bibnamefont{Dmitra{\v s}inovi{\' c}}},
  \bibinfo{journal}{Phys. Rev. D} \textbf{\bibinfo{volume}{57}},
  \bibinfo{pages}{7019} (\bibinfo{year}{1998}).

\bibitem{nambu61}
\bibinfo{author}{\bibfnamefont{Y.}~\bibnamefont{Nambu}} \bibnamefont{and}
  \bibinfo{author}{\bibfnamefont{G.}~\bibnamefont{Jona-Lasinio}},
  \bibinfo{journal}{Phys. Rev.} \textbf{\bibinfo{volume}{122}},
  \bibinfo{pages}{345} (\bibinfo{year}{1961}).

\bibitem{kars00}
\bibinfo{author}{\bibfnamefont{F.}~\bibnamefont{Karsch}},
  \bibinfo{journal}{(Proc. Suppl.) Nucl. Phys. B}
  \textbf{\bibinfo{volume}{83--84}}, \bibinfo{pages}{14}
  (\bibinfo{year}{2000}).

\bibitem{bonn96}
\bibinfo{author}{\bibfnamefont{C.}~\bibnamefont{Ritter}},
  \bibinfo{author}{\bibfnamefont{B.~C.} \bibnamefont{Metsch}},
  \bibinfo{author}{\bibfnamefont{C.~R.} \bibnamefont{M{\" u}nz}},
  \bibnamefont{and} \bibinfo{author}{\bibfnamefont{H.~R.} \bibnamefont{Petry}},
  \bibinfo{journal}{Phys. Lett. B} \textbf{\bibinfo{volume}{380}},
  \bibinfo{pages}{431} (\bibinfo{year}{1996}).

\bibitem{dm01}
\bibinfo{author}{\bibfnamefont{V.}~\bibnamefont{Dmitra{\v s}inovi{\' c}}},
  \bibinfo{journal}{Nucl. Phys. A} \textbf{\bibinfo{volume}{686}},
  \bibinfo{pages}{379} (\bibinfo{year}{2001}).

\bibitem{wein67}
\bibinfo{author}{\bibfnamefont{S.}~\bibnamefont{Weinberg}},
  \bibinfo{journal}{Phys. Rev. Lett.} \textbf{\bibinfo{volume}{18}},
  \bibinfo{pages}{507} (\bibinfo{year}{1967}).

\bibitem{dmit96}
\bibinfo{author}{\bibfnamefont{V.}~\bibnamefont{Dmitra{\v s}inovi{\' c}}},
  \bibinfo{author}{\bibfnamefont{S.}~\bibnamefont{Klevansky}},
  \bibnamefont{and} \bibinfo{author}{\bibfnamefont{R.~H.}
  \bibnamefont{Lemmer}}, \bibinfo{journal}{Phys. Lett. B}
  \textbf{\bibinfo{volume}{386}}, \bibinfo{pages}{45} (\bibinfo{year}{1996}).

\bibitem{nieh67}
\bibinfo{author}{\bibfnamefont{H.~T.} \bibnamefont{Nieh}},
  \bibinfo{journal}{Phys. Rev.} \textbf{\bibinfo{volume}{163}},
  \bibinfo{pages}{1769} (\bibinfo{year}{1967}).

\bibitem{fgz64}
\bibinfo{author}{\bibfnamefont{R.~P.} \bibnamefont{Feynman}},
  \bibinfo{author}{\bibfnamefont{M.}~\bibnamefont{Gell-Mann}},
  \bibnamefont{and} \bibinfo{author}{\bibfnamefont{G.}~\bibnamefont{Zweig}},
  \bibinfo{journal}{Phys. Rev. Lett.} \textbf{\bibinfo{volume}{13}},
  \bibinfo{pages}{678} (\bibinfo{year}{1964}).

\bibitem{pasc82}
\bibinfo{author}{\bibfnamefont{P.}~\bibnamefont{Pascual}} \bibnamefont{and}
  \bibinfo{author}{\bibfnamefont{E.}~\bibnamefont{de~Rafael}},
  \bibinfo{journal}{Z. Phys. C} \textbf{\bibinfo{volume}{12}},
  \bibinfo{pages}{127} (\bibinfo{year}{1982}).

\bibitem{hatsuda01}
\bibinfo{author}{\bibfnamefont{T.}~\bibnamefont{Hatsuda}}, in
  \emph{\bibinfo{booktitle}{Hadrons and Nuclei}}, edited by
  \bibinfo{editor}{\bibfnamefont{S.-W.} \bibnamefont{Hong}}
  (\bibinfo{publisher}{AIP}, \bibinfo{address}{Melville, NY},
  \bibinfo{year}{2001}).

\end{thebibliography}
\end{document}